\documentstyle[aps,12pt,epsf]{revtex}
\begin{document}
\begin{center}
{\large \bf Doubly Charmed Baryon Production in Hadronic
Experiments}\\
\vspace*{5mm}
Berezhnoy A.V., \underline{Kiselev V.V.}, 
Likhoded A.K., Onishchenko A.I. \\
{\sf State Research Center of Russia "Institute for High Energy
Physics"} \\
{\it Protvino, Moscow region, 142284 Russia}\\
Fax: +7-095-2302337\\
E-mail: kiselev@mx.ihep.su
\end{center}
\centerline{Abstract}
\abstract{
In the leading order of perturbative QCD one calculates the total and
differential cross-sections for the hadronic production of
doubly charmed baryons $\Xi_{cc}$ and $\Xi_{cc}^*$  in different
experiments. The experimental evaluation of cross-sections for the
$J/\Psi +D + \bar D$ production would allow one to decrease the
uncertainty in the determination of cross-sections for the
doubly charmed baryons due to the choice of $\alpha_s$ and $m_c$.
One shows that in the HERA-B and E781 experiments with fixed tagets
the suppression of the $\Xi_{cc}$ and $\Xi_{cc}^*$ production
to the yield of $c \bar c$-pairs is the value of the order of
$10^{-6}-10^{-5}$, whereas at the TEVATRON and LHC colliders
it is about $10^{-4}-10^{-3}$. In the E781 experiment
the observation of $\Xi_{cc}$ and $\Xi_{cc}^*$ is practically
unpossible.
At the HERA-B and TEVATRON facilities one can
expect $10^5$ events with the double charm, and at LHC one has about
$10^9$ ones.
}\\

\vspace*{14mm}
PACS numbers: 14.20.Lq, 13.60.Rj, 12.38.Bx, 13.30.-a
\newpage
\section{Introduction}
Recent years  are marked by a rapid increase of charmed particles
observed in modern experiments. So, the study of about $10^6$
charmed particles is expected at fixed target
FNAL facilities of E831 and E781.
An increase of this value by two orders of magnitude is proposed in
experiments of next generation. Along with standard problems of
CP-violation
in the charmed quark sector and a measuring of rare decays etc.,
an investigation of processes with more than one $c\bar c$-pair
production becomes actual. The production of  additional $c\bar
c$-pair
strongly decreases a value of  cross-section for  such processes.
This fact must be especially taken into account in fixed
target experiments, where the quark-partonic luminosities are
strongly suppressed in the region of  heavy mass production.

An interesting process of the mentioned kind is the doubly charmed
baryon
production. The doubly charmed $\Xi_{cc}^{(*)}$-baryon
represents an absolutely  new type of
objects in comparison with the ordinary baryons containing
light quarks only. The basic state of  such baryon  is analogous
to a $(\bar Q q)$-meson, which contains one heavy antiquark
$\bar Q$ and one light quark $q$. In the doubly heavy baryon the role
of heavy antiquark is played by the $(cc)$-diquark, which is in
antitriplet color-state \cite{1}. It has a small size in comparison
with the scale of
the light quark confinement.

The spectrum of  $(ccq)$-system states has to  differ
essentially from the heavy meson
spectra, because the composed $(cc)$-diquark has a set of the excited
states (for example, $2S$ and $2P$) in contrast to the heavy quark.
The
energy of diquark excitation is twice less than the excitation
energy of light quark bound with the diquark. So, the
representation on the
compact diquark can be straightforwardly connected with the level
structure
of  doubly heavy baryon%
\footnote{Our estimates  of diquark mass in the Martin \cite{2}
potential with taking into account the color factor for antitriplet
state
of the quark pair and using the results of heavy quark effective
theory
\cite{3}, give
the value of $M(\Xi_{cc}^{(*)})=3.615\pm 0.035$ GeV
(without taking into account a spin  dependent  interaction).
The  mass
shift of  vector diquark is determined by the formula
$\delta M\simeq \frac{1}{2}|R_{cc}(0)|^2/
|R_\psi(0)|^2 (M_\psi-M_{\eta_c})/4\simeq 5$ MeV.
The splitting between $\Xi_{cc}$ and $\Xi_{cc}^*$ is equal to
 $\Delta M(\Xi_{cc}^{(*)})\simeq   \frac{3}{4}\Delta
 M(D^{(*)})\simeq
108$ MeV, so
$M(\Xi_{cc})=3.584\pm 0.035$ GeV, $M(\Xi_{cc}^*)=3.638\pm 0.035$ GeV.
The diquark size $r_{cc}\sim 0.5$ fm is close to that
of $J/\Psi$.}.

Another interesting aspect of the doubly charmed baryon researches is
a production mechanism. The $(ccq)$-baryon production was discussed
in the
number of papers \cite{4}-\cite{7}. The main problem of
calculations is reduced to an evaluation of
the production cross section for the diquark in the antitriplet color
state.
One assumes further that the $(cc)$-diquark nonperturbatively
transforms
into the $(ccq)$-baryon with a probability close to unit.
The hadronic production of diquark is subdivided into two parts.
The first stage is the hard production of two $(c\bar c)$-pairs  in
the
processes of $gg \to c \bar c c \bar c$ and $q \bar q \to c \bar c c
\bar c$,
which are described by the Feynman diagrams of the fourth order over
the $\alpha_s$ coupling constant \cite{bps}.  The second step is the
nonperturbative
fusion of two $c$-quarks  with a small relative momentum into
the $(cc)$-diquark. For the $S$-wave states, this process is
characterized
by  the radial wave function at the origin, $R(0)$.

The main difference between  the existing evaluations of the doubly
charmed
baryon cross section consists in the methods used for  the hard
subprocess calculation.
 In  paper \cite{8} a part of diagrams connected with the
$c$-fragmentation  into the $(c c)$-diquark 
is only used instead of the complete set of diagrams.
As was shown in paper \cite{6}
this estimation is not absolutely correct, because it becomes true
only at $p_T> 35$ GeV, where the fragmentation mechanism is dominant.
In other kinematical regions the application of
fragmentational approximation is not justified and it leads
to wrong results, especially at $\sqrt{\hat s}$ being not much
greater than  $p_T^{min}$. 

In the framework
of fragmentation model the account for the leading logarithm corrections
can be performed in a simple way. However, in the framework
of calculations for the complete set  of diagrams in the
$\alpha_s^4$-order the regime of fragmetation is dominant only in the restricted region
of kinematical variables, $p_T>p_T^{min}$, $s>s^{min}$, wherein the
procedure for the taking into account the evolution of fragmentation
functions
does not cause some difficulties. But the account for the emission of
additional hard gluon beyond the mentioned region requires the
calculations
in the next $\alpha_s^5$-order of perturbative QCD, so that the
number of
diagrams increases drastically, and there are no large logs like
$\log(p_T^2/m^2)$ or $\log(s/m^2)$ in the dominant kinematical domain.
So, following refs.\cite{4}-\cite{7}, in this paper we are restricted by the Born
approximation.

However, even after taking into account the
complete  set of diagrams  essential   uncertainties  in the
estimations
of the $(ccq)$-baryon production remain. The basic parameters
determining these
uncertainties are the values of $\alpha_s$, $m_c$ and $R_{cc}(0)$.
 In addition, it is not clear, to what extent the hypothesis on the
 hadronization
of $(cc)$-diquark into the $(ccq)$-baryon with the unit probability
is correct
or not. The matter is that the interaction between the diquark and
gluons
is not suppressed in contrast to the $(c \bar c)$-pair
production in the color singlet state,
when the quarkonium dissociation supposes exchange with
the quark-gluonic sea by  two hard gluons with virtualities,
which are greater than the inverse size of quarkonium.

A decrease of the uncertainty in the $(ccq)$-baryon cross-section
would be possible by means of comparing
the process of baryon production with the
analogous process of $J/\Psi + D \bar D$ production\footnote{
We calculate the $J/\psi+c\bar c$
production and
assume that the $c\bar c$-pair transforms to $D\bar D+$ some light
hadrons with the probability very close to unit. So, we neglect
the production of charmed baryons as well as bound states of
charmonium.}. The latter is
described by practically the same diagrams of fourth order with the
well-known
wave function of $J/\Psi$ at the origin%
\footnote{The value of $|R_{\Psi}(0)|$ is determined by the width of
leptonic decay, $J/\Psi\to l^+l^-$ with taking into account the hard
gluonic correction, so, numerically,
$|R_{\psi}(0)|=\sqrt{\pi M/3}\tilde f_{\psi}$, where
$\tilde f_{\psi}=540$ MeV.}.

In this way of connection to the $J/\Psi+D\bar D$ process one could
remove
the part of  uncertainties, which are due to  $\alpha_s$ and $m_c$
in the $(cc)$-diquark production process.
    In the following sections of the paper the joint cross-section
    calculations
of these processes in $\pi^-p$ and $pp$ interactions are performed.

There are  some papers, wherein the fragmentation model
for the
$J/\psi$ production at $p_T>5$ GeV was constructed with the account
for the color-octet contribution (see the review \cite{bfy}). 
However, the associated production of
$J/\psi+D\bar D$
requires the production of additional $c\bar c$-pair, so that the
production
mechanism is evidently the other one, and the $J/\psi+D\bar D$ yield
is only a small fraction of the inclusive $J/\psi$ production
in $pp$-interactions.

For the $B_c+b\bar c$ production one has shown that the regime of
fragmentation becomes dominant at $p_T>35$ GeV. The production
of $J/\psi+c\bar c$ is completely analogous to the associated
production
of $B_c+b\bar c$ (with the careful account for the identity of
charmed
quarks). For the $B_c+b\bar c$, $\Xi_{cc}+\bar c\bar c$ and
$J/\psi+c\bar c$ production one finds the common regularity:
the fragmentation regime is displaced to the region of $p_T>35$ GeV.
To convince completely, we present the corresponding figure for the
gluon-gluon subprocess of $gg\to J/\psi+c\bar c$ at $\sqrt{\hat
s}=100$ GeV. Thus, we insist on the statement that for the associated
production of
$J/\psi+c\bar c$ and $\Xi_{cc}+\bar c \bar c$ the fragmentation works
at $p_T\gg m_c$.

Section II is devoted to the description of production models for
the $(ccq)$-baryons and $J/\Psi +D \bar D$. In Section III one
presents
the calculation results for the production  cross-section of 
$(ccq)$-baryons and $J/\Psi +D \bar D$ in the
fixed target experiments E781 and HERA-B.

\section{Production Mechanism}
As was mentioned in Introduction, we suppose that the diquark
production can be subdivided by two stages. On the first stage the
production amplitude of four free quarks is calculated for the
following processes
\begin{eqnarray}
&& gg \to cc \bar c \bar c, \\
&& q\bar q \to cc \bar c \bar c.
\end{eqnarray}

The calculation technique applied in this work is analogous to that
for the hadronic production of $B_c$ \cite{9}, but in this case the
bound
state is composed by two quarks \cite{5,6} instead of the
quark and antiquark.

One assumes that the binding energy in the diquark is much less than
the
masses of constituent quarks and, therefore, these quarks
are on the mass shells.  So, the quark four-momenta  are related
to the $(Q_1 Q_2)$ diquark momentum in the following way
\begin{equation}
p_{Q_1}=\frac{m_{Q_1}}{M_{(Q_1 Q_2)}}P_{(Q_1 Q_2)}
\;,\;\; \quad p_{Q_2}=\frac{m_{Q_2}}{M_{(Q_1 Q_2)}}P_{(Q_1 Q_2)},
\end{equation}
where  $M_{(Q_1 Q_2)}=m_{Q_1}+m_{Q_2}$ is the diquark mass,
$m_{Q_1}, m_{Q_2}$ are the quark masses.

In the given approach the diquark production is described by
36 Feynman diagrams of the leading order, corresponding to the
production
of four free quarks with the combining of two quarks into the color
antitriplet
diquark with the given quantum numbers over the Lorentz group. The
latter
procedure is performed by means of the projection operators
\begin{equation}
\displaystyle
{\cal N}(0,0)=
\sqrt{\frac{2M_{(Q_1 Q_2)}}{2m_{Q_1}2m_{Q_2}}}
\frac{1}{\sqrt{2}}\{ \bar u_1(p_{Q_1},+)\bar u_2(p_{Q_2},-)
- \bar u_1(p_{Q_1},-)\bar u_2(p_{Q_2},+) \},
\end{equation}
for the scalar state of diquark (the
corresponding baryon is denoted as $\Xi_{Q_1 Q_2}'(J =1/2)$);
\begin{eqnarray}
\displaystyle
{\cal N}(1,-1) &=&
\sqrt{\frac{2M_{(Q_1 Q_2)}}{2m_{Q_1}2m_{Q_2}}}
\bar u_1(p_{Q_1},-)\bar u_2(p_{Q_2},-), \nonumber\\
{\cal N}(1,0) &=&
\sqrt{\frac{2M_{(Q_1 Q_2)}}{2m_{Q_1}2m_{Q_2}}}
\frac{1}{\sqrt{2}}\{ \bar u_1(p_{Q_1},+)\bar u_2(p_{Q_2},-) +
\bar u_1(p_{Q_1},-)\bar u_2(p_{Q_2},+) \}, \nonumber\\
{\cal N}(1,+1) &=&
\sqrt{\frac{2M_{(Q_1 Q_2)}}{2m_{Q_1}2m_{Q_2}}}
\bar u_1(p_{Q_1},+)\bar u_2(p_{Q_2},+) 
\end{eqnarray}
for the vector state of diquark
(the baryons are denoted as  $\Xi_{Q_1 Q_2}(J =1/2)$ and
$\Xi_{Q_1 Q_2}^*(J =3/2)$).

To produce the quarks, composing the diquark in the $\bar 3_c$ state,
one has to introduce the color wave function as
$\varepsilon_{ijk}/\sqrt{2}$, into the diquark production vertex,
so that $i=1,2,3$ is the color index of the first quark, $j$ is that
of the second one, and $k$ is the color index of diquark.

The diquark production amplitude $A_k^{S s_z}$ is expressed through
the
amplitude $T_k^{S s_z}(p_i)$ for the free quark production in
kinematics
(3) 
\begin{equation}
A^{Ss_z}_k=\frac{R_{Q_1 Q_2}(0)}{\sqrt{4\pi}}T^{Ss_z}_k(p_i),
\end{equation}
where $R_{Q_1 Q_2}(0)$ is the diquark radial wave function at the
origin,
$k$ is the color state of diquark, $S$ and $s_z$ are the diquark spin
and diquark spin projection on the $z$-axis, correspondingly.

In the numerical calculation giving the results, which will be
discussed in the next Section, one supposes the following values of
 parameters
\begin{eqnarray}
&&\alpha_s=0.2,  \nonumber\\
&&m_c=1.7\; {\rm GeV}, \nonumber\\
&&R_{cc(1S)}(0)=0.601\; {\rm GeV}^{3/2},
\end{eqnarray}
where the value of $R_{cc}(0)$ has been calculated by means of
numerical
solution of the Schr\"odinger equation with the Martin potential
\cite{10},
multiplied by the 1/2 factor caused by the color antitriplet state of
quarks instead of the singlet one.

To calculate the production cross-section of diquarks composed of
two $c$-quarks, one has to account for their identity. One can easily
find, that the antisymmetrization  over the identical fermions leads
to
the scalar diquark amplitude equal to zero, and it results in the
amplitude of the vector $(cc)$-diquark production being obtained
by the substituting of equal masses in the production amplitude of
vector
diquark composed of two quarks with the different flavors, and taking
into account the 1/2 factor for the identical quarks and antiquarks.

In this work one supposes that the produced diquark forms the baryon
with the unit probability by catching up the light quark from
the quark-antiquark sea at small $p_T$ or having the fragmentation
into
the baryon at large $p_T$.

\begin{figure}[p]
\hspace*{-1.75cm}
\epsfxsize=14cm \epsfbox{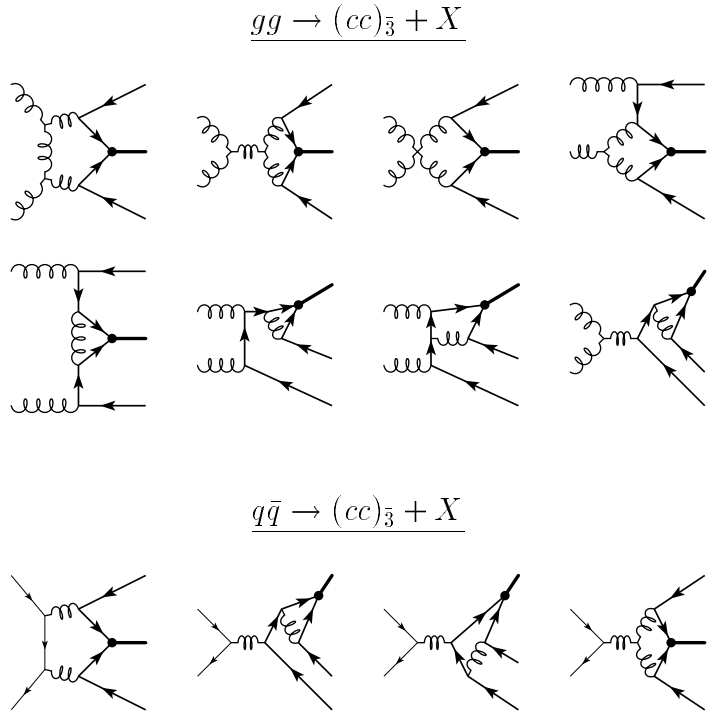}
\vspace*{-8cm}
\caption{
The examples of diagrams  for the gluon-gluon and
quark-antiquark production of $(cc)$-diquark.
The initial quarks are denoted by the thin  fermion lines,
the final quarks are denoted by the bold fermion lines and
the gluons are denoted by the helical lines.}
\label{fig1}
\end{figure}

The typical diagrams of fourth order describing  the processes (1)
and (2) with the binding of $cc$-pair into the diquark
are shown in Fig. \ref{fig1}. One can subdivide them into two groups.
The first
group contains the diagrams of fragmentation type, wherein the
$(c\bar c)$-pair emits another one.
The second group corresponds to the independent dissociation of
gluons
into the $(c\bar c)$-pairs with the following fusion into the
diquark.
The diagrams of second group belong to the recombination type.  
As was mentioned above,
the authors of some papers restricted themselves by the consideration
of
fragmentation diagrams, only. In this way they reduced the
cross-section formulae to the $(c\bar c)$-pair production
cross-section
multiplied  by the fragmentation function of $c$-quark into the
$(cc)$-diquark.
As was shown in \cite{6}, the latter approach is correct only under
the two
following conditions: $M_{(cc)}^2\ll \hat s$ and $p_T \gg M_{(cc)}$.
In other kinematical regions, the contribution of recombination
diagrams
dominates.

The typical value of $p_T$, wherefrom the fragmentation begins to
dominate, is
$p_T>35$ GeV. It is clear, that at realistic $p_T$ one has to take
into account all contributions including the recombination one.
For the first time, the complete set of diagrams was
taken into account in \cite{6} and, after that, in
\cite{7}.  In the both papers the calculations are performed 
only for the gluon-gluon
production, which is a rather good approximation at collider
energies. 
For the fixed target experiments the value  of total  energy
strongly decreases,
and, hence, the values of energy in the subprocesses (1)
and (2) decrease too.

The contribution of  quark-antiquark annihilation becomes essential
at
fixed target energies, especially for the processes
with initial valent antiquarks. In the following consideration one
allows
for the quark-antiquark annihilation into  four free
charmed quarks in the estimation of
yield for the doubly charmed baryon. To our knowledge, the
corresponding calculations for the diquark-production
were not yet performed early, so, they are carried out
here for the first time.

\section{ Doubly  Charmed Baryon Production
in Fixed Target Experiments }

The applied method of calculations is 
the same as in our previous works \cite{6,9}.
One calculates the complete set of diagrams in the fourth order 
over the strong coupling
constant for the Born amplitude of process under consideration.

\begin{figure}[p]
\vspace*{1.cm}
\hbox to 1.5cm {\hfil\mbox{$\hat \sigma_{gg}^{cc}$,
 $\hat \sigma_{gg}^{J/\Psi+D\bar D}$, pb}}
\epsfxsize=14cm \epsfbox{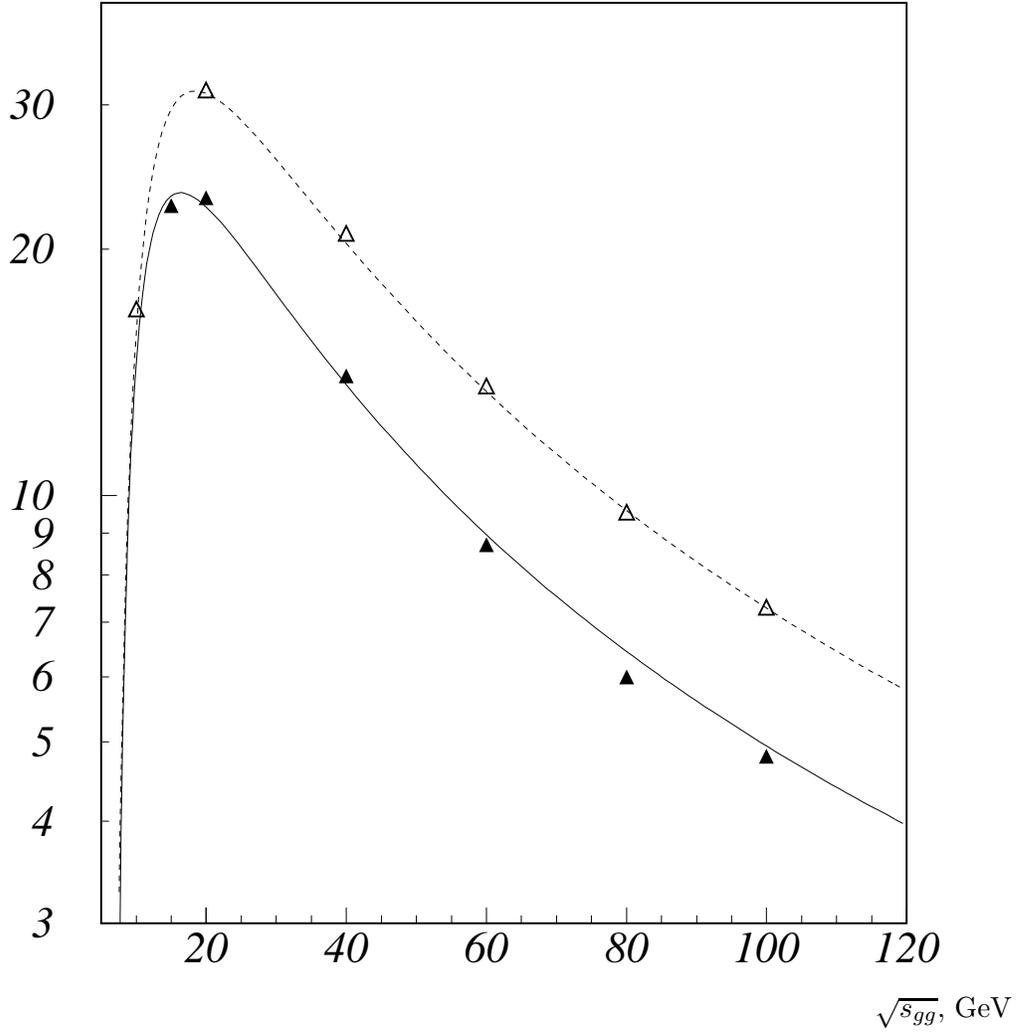}

\vspace*{-1.cm}
\hbox to 14.cm {\hfil \mbox{$\sqrt{s_{gg}}$, GeV}}
\vspace*{1.cm}
\caption{
The total  cross-section of the gluon-gluon production of
$(cc)$-diquark
(solid triangle) and  $J/\Psi+D\bar D$ (empty triangle)
in comparison with the approximations of (8) and (10) (solid and
dashed curves, correspondingly).}
\label{fig2}
\end{figure}

\begin{figure}[p]
\vspace*{1.cm}
\hbox to 1.5cm {\hfil\mbox{$\hat \sigma_{q \bar q}^{cc}$,
 $\hat \sigma_{q \bar q}^{J/\Psi+D\bar D}$, pb}}
\epsfxsize=14cm \epsfbox{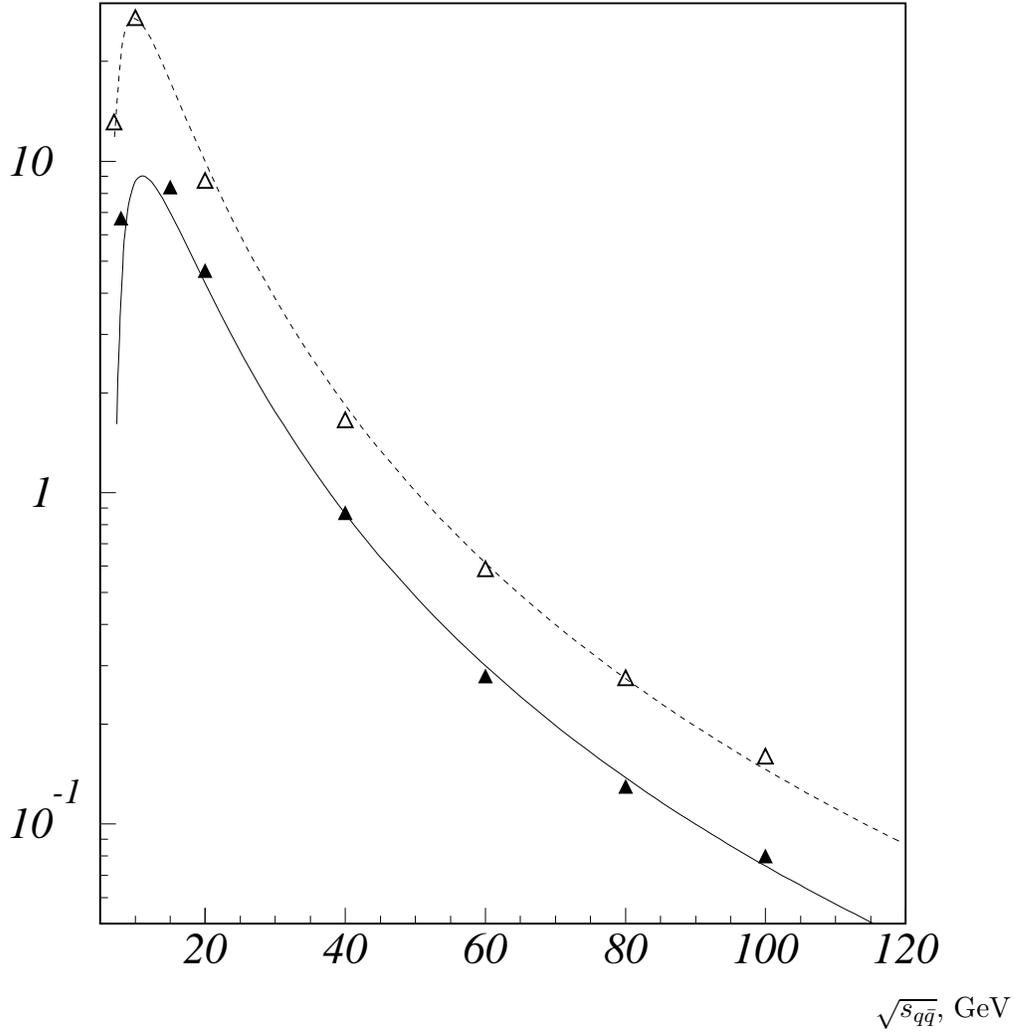}

\vspace*{-1.cm}
\hbox to 14.cm {\hfil \mbox{$\sqrt{s_{q \bar q}}$, GeV}}
\vspace*{1.cm}
\caption{
The total  cross-section of the quark-antiquark production of
$(cc)$-diquark
(solid triangle) and  $J/\Psi+D\bar D$ (empty triangle)
in comparison with the approximations of (9) and (11) (solid and
dashed curves, correspondingly).}
\label{fig3}
\end{figure}

The calculation results for the total cross-section of the diquark-production
subprocesses versus the total energy are shown in Figs. \ref{fig2} and
\ref{fig3} for the given
values of $\alpha_s$, $m_c$ and $R_{(cc)}(0)$. These dependencies can
be approximately described by the following expressions
\begin{eqnarray}
&& \hat\sigma_{gg}^{(cc)}=213.\left (1-\frac{4m_c}{\sqrt{\hat s}}
\right )^{1.9} \left (\frac{4m_c}
{\sqrt{\hat s}}\right )^{1.35}\quad {\rm pb},\\
&& \hat\sigma_{q\bar q}^{(cc)}=
206.\left (1-\frac{4m_c}{\sqrt{\hat s}}\right )^{1.8}
\left (\frac{4m_c}
{\sqrt{\hat s}} \right )^{2.9}\quad {\rm pb }.
\end{eqnarray}

One has to mention that the numerical coefficients  depend on the
model
parameters, so that $\hat \sigma\sim \alpha_s^4|R(0)|^2/m_c^5$.

As was mentioned in the Introduction, the production of $J/\Psi$ in
the
subprocesses of $gg \to J/\Psi + c\bar c$ and $q\bar q \to J/\Psi
+c\bar c$
is also calculated in this work. The numerical results of such
consideration
are shown  in Figs. \ref{fig2} and \ref{fig3}. The parameterization
of these results
versus the energy $\sqrt{\hat s }$ are presented below
\begin{eqnarray}
&& \hat\sigma_{gg}^{J/\Psi}=518.\left (1-
\frac{4m_c}{\sqrt{\hat s}}\right )^{3.0} \left (\frac{4m_c}
{\sqrt{\hat s}} \right )^{1.45}\quad {\rm pb}, \\
&& \hat\sigma_{q\bar q}^{J/\Psi}=699.\left( 1- 
\frac{4m_c}{\sqrt{\hat s}}\right )^{1.9}
\left ( \frac{4m_c}{\sqrt{\hat s}} \right )^{2.97}\quad {\rm pb}.
\end{eqnarray}

\begin{figure}[p]
\hbox to 1.5cm {\hfil\mbox{$d\hat \sigma_{gg}^{J/\Psi+c\bar
c}/dp_T$, pb/GeV}}
\epsfxsize=14cm \epsfbox{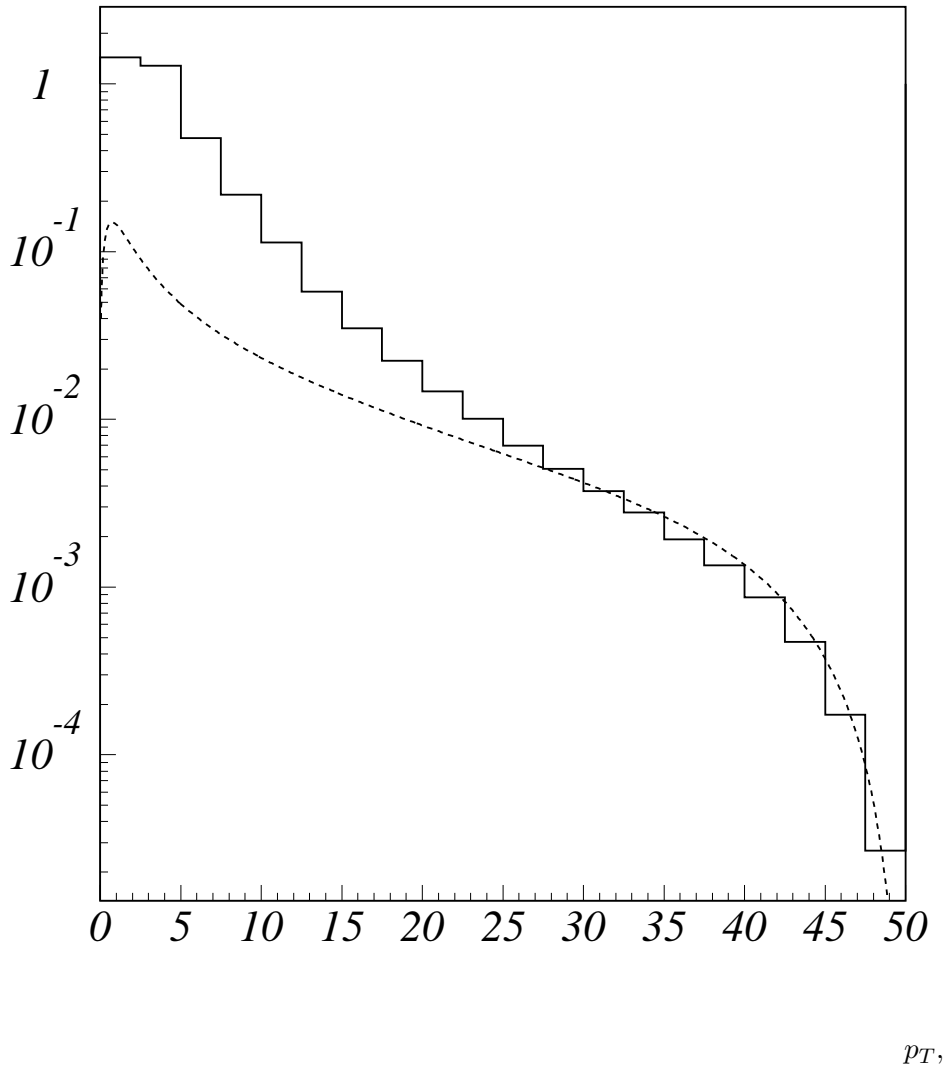}

\vspace*{-0.5cm}
\hbox to 14.cm {\hfil \mbox{$p_T$, GeV}}
\vspace*{1.cm}
\caption{
The differential cross-section for the associated production of
$J/\Psi+c\bar c$  in the gluon-gluon subprocess at 100 GeV (solid
histogram) in comparison with the prediction of fragmentation model
(dashed curve), correspondingly.}
\label{add}
\end{figure}

As well as for the $B_c+b\bar c$, $\Xi_{cc}+\bar c\bar c$ production,
one finds the following regularity for the $J/\psi+c\bar c$ production:
the fragmentation regime is displaced to the region of $p_T>35$ GeV.
The latter fact can be certainly observed in the figure for the
gluon-gluon subprocess of $gg\to J/\psi+c\bar c$ at $\sqrt{\hat
s}=100$ GeV (Fig. \ref{add}). Thus,  for the associated
production of $J/\psi+c\bar c$ and $\Xi_{cc}+\bar c \bar c$ 
the fragmentation works at $p_T\gg m_c$.

These formulae quite accurately 
reconstruct the results of precise calculations at
$\sqrt{\hat s} < 150$ GeV, and that is why they can be used for the
approximate estimation of total hadronic production cross-section
for the $(cc)$-diquark and $J/\Psi$  by means of their convolution 
with the partonic distributions as below
\begin{equation}
\sigma=\sum_{i,j}\int dx_1 dx_2 f_{i/A}(x_1,\mu) f_{j/B}(x_2,\mu)
\hat\sigma ,
\end{equation}
where  $f_{i/A}(x,\mu)$ is the distribution of $i$-kind parton
in the  $A$-hadron.
The parton distributions  used for the proton are the CTEQ4 
parameterizations \cite{10}, and
those of used for the $\pi^-$-meson are the Hpdf ones \cite{11}. In
both these
cases the virtuality scale is fixed at 10 GeV. 
As for the choice of fixed scale in the structure functions,
this is caused by the fact that the cross-section of subprocesses
is integrated in the region of low $\hat s$ close to the fixed scale,
so that the account of "running" scale weakly  changes the estimate of
$\Xi_{cc}$-baryon yield in comparison with the mentioned uncertainty
of diquark model (we have found the scale-dependent variation to be
at the level of $\delta \sigma/\sigma \sim 10$ \%).

\begin{figure}[p]
\hbox to 1.5cm {\hfil\mbox{$ \sigma_{\pi^- p}^{cc}$,
 $\hat \sigma_{\pi^- p}^{J/\Psi+D\bar D}$, nb}}
\epsfxsize=14cm \epsfbox{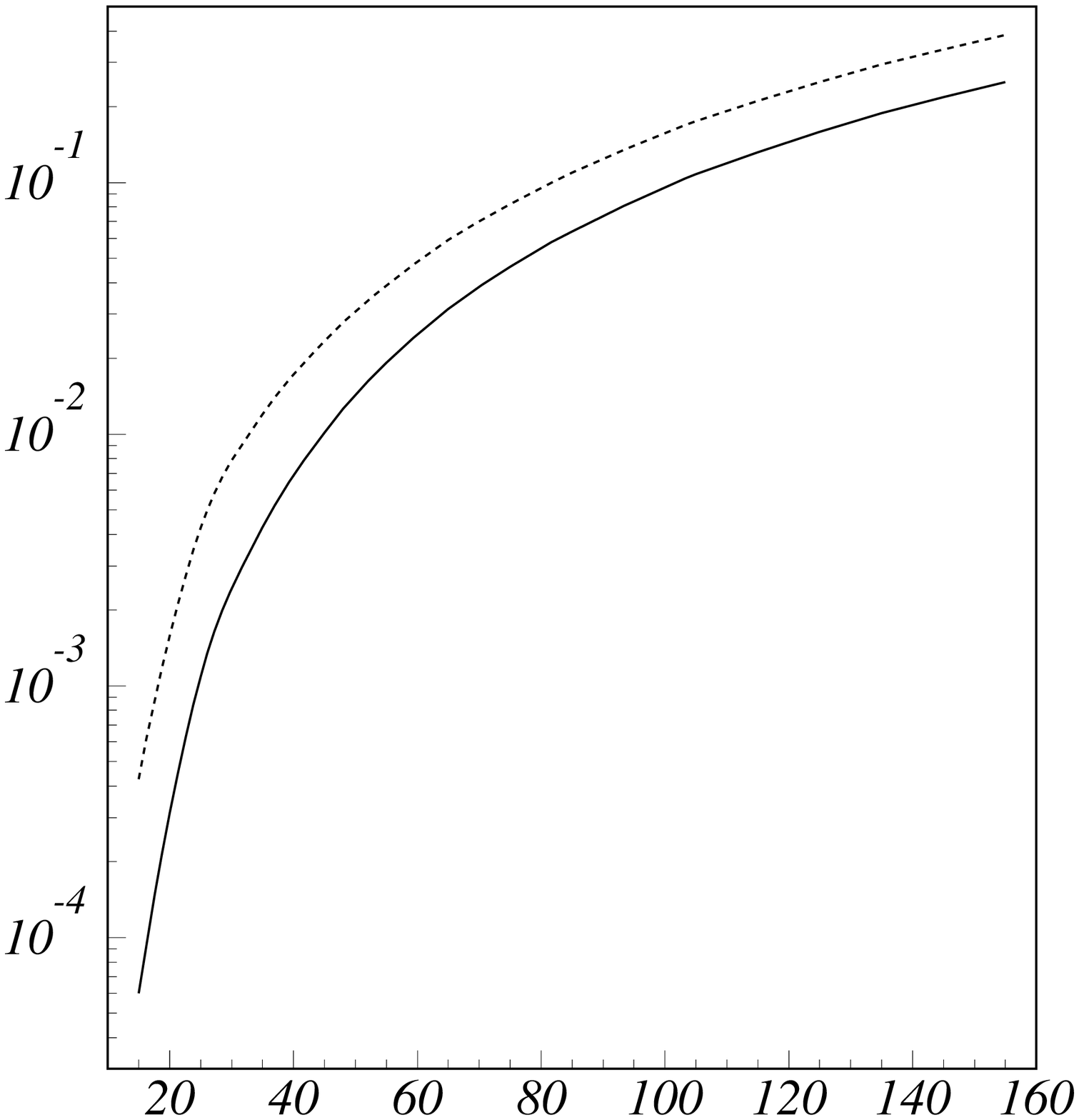}

\hbox to 14.cm {\hfil \mbox{$\sqrt{s_{\pi^- p}}$, GeV}}
\vspace*{1.cm}
\caption{
The total  cross-section of the pion-proton production of
$(cc)$-diquark
and  $J/\Psi+D\bar D$  (solid and
dashed curves, correspondingly).}
\label{fig4a}
\end{figure}

\begin{figure}[p]
\hbox to 1.5cm {\hfil\mbox{$ \sigma_{p p}^{cc}$,
 $\hat \sigma_{p p}^{J/\Psi+D\bar D}$, nb}}
\epsfxsize=14cm \epsfbox{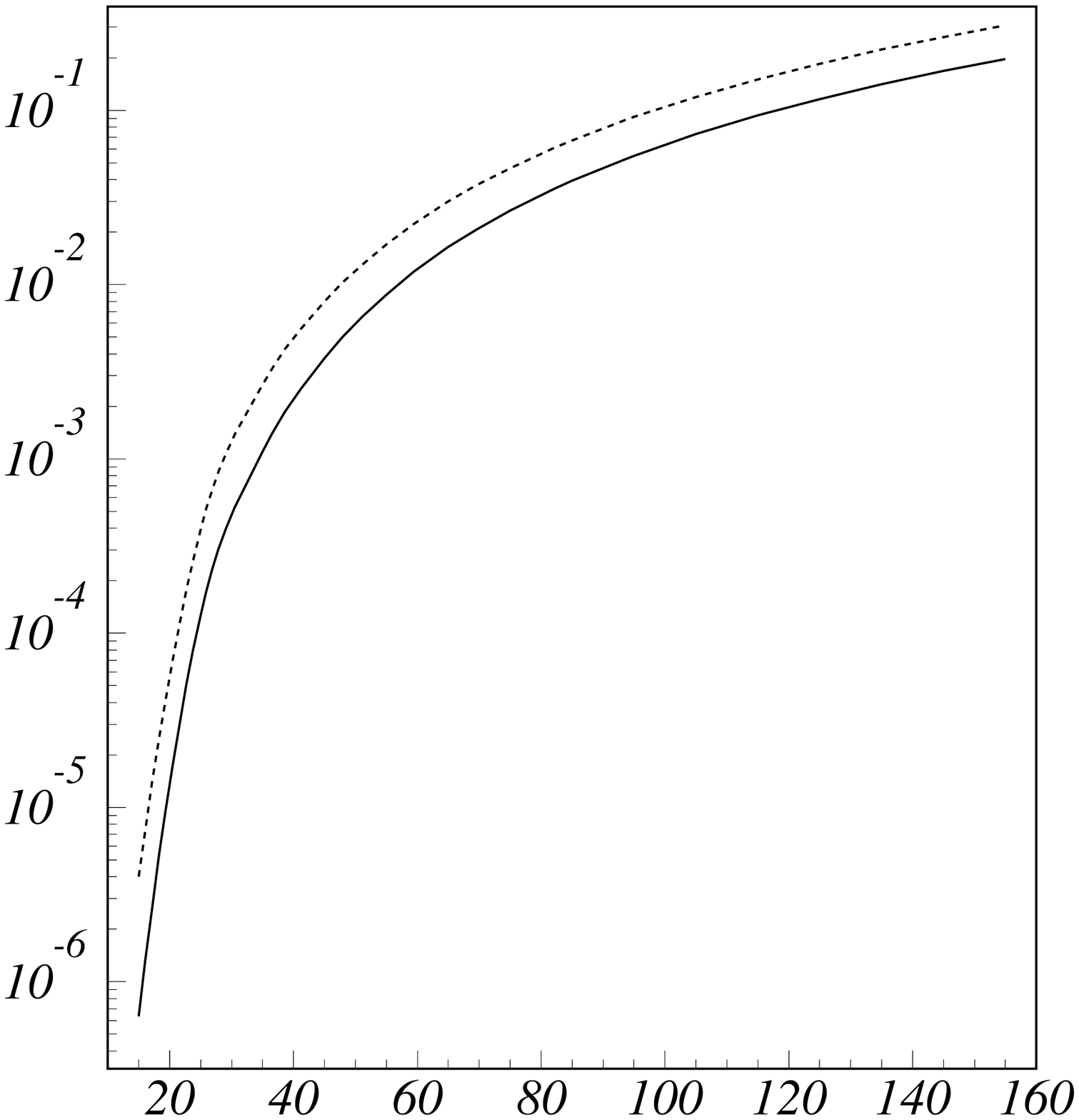}

\hbox to 14.cm {\hfil \mbox{$\sqrt{s_{p p}}$, GeV}}
\vspace*{1.cm}
\caption{
The total  cross-section of the proton-proton production of
$(cc)$-diquark
and  $J/\Psi+D\bar D$  (solid and
dashed curves, correspondingly).}
\label{fig4b}
\end{figure}

The total hadronic
production cross-section for these processes are presented in Figs.
\ref{fig4a}
and \ref{fig4b} for the $\pi^-p$ and  $pp$-interactions,
correspondingly.  As one can see in Figs. \ref{fig4a} and
\ref{fig4b}, 
the cross-section
of $(cc)$-diquark as well as the cross-section of $J/\Psi+c\bar c$
are strongly suppressed at low energies in comparison with the values
at
the collider energies.

\begin{figure}
\hbox to 1.5cm {\hfil\mbox{
$d\sigma^{cc}_{\pi^- p}/d p_T$, 
 $d\sigma^{J/\Psi+D\bar D}_{\pi^- p}/d p_T$, pb/GeV}}
\epsfxsize=14cm \epsfbox{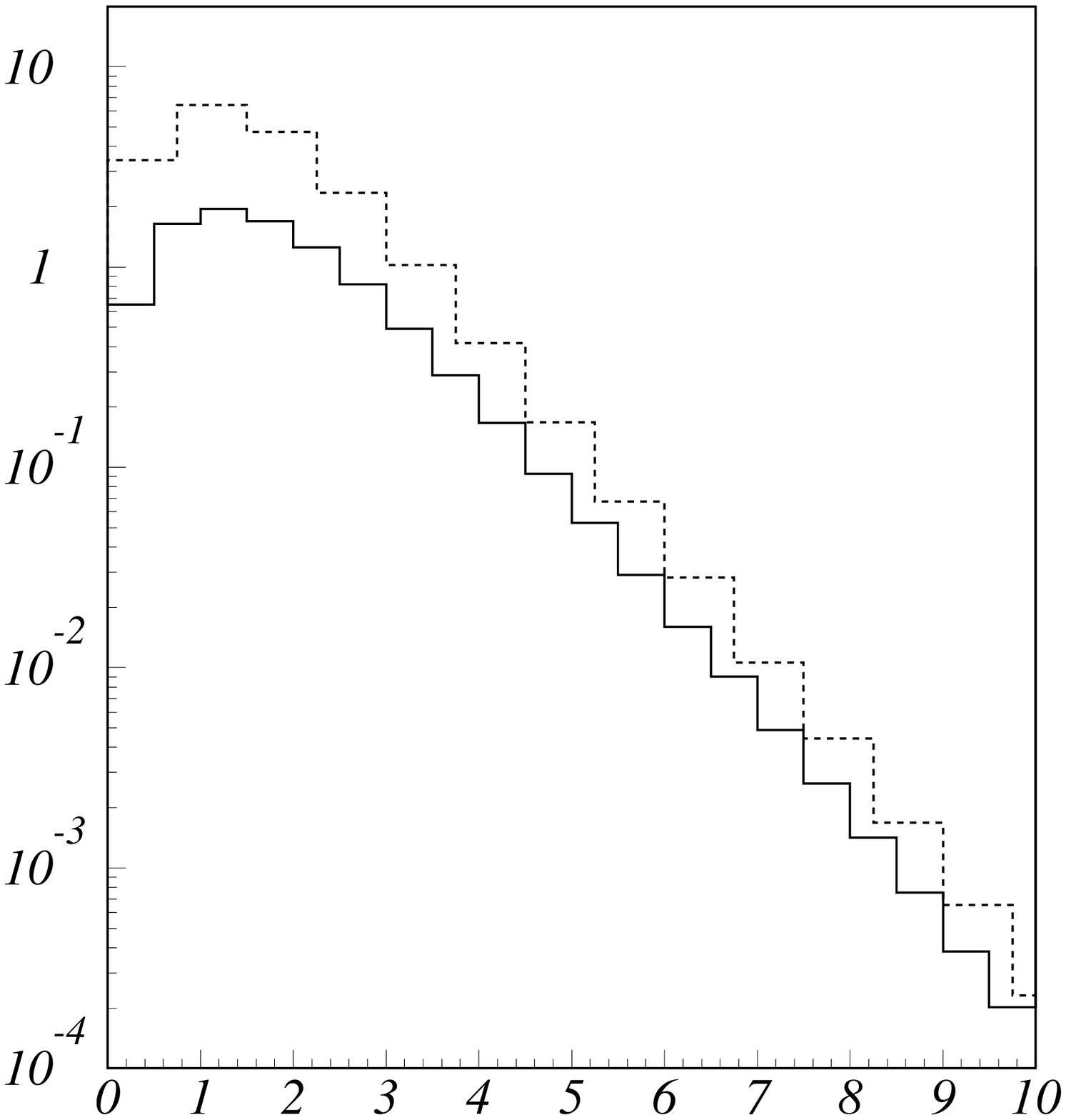}

\vspace*{-1.cm}
\hbox to 14.cm {\hfil \mbox{$p_T$, GeV}}
\vspace*{1.cm}
\caption{
$d\sigma^{cc}_{\pi^- p}/d p_T$ (solid histogram)
and $d\sigma^{J/\Psi+D\bar D}_{\pi^- p}/d p_T$ (dashed histogram)
at the pion-proton interaction energy of 35 GeV.}
\label{fig5a}
\end{figure}

\begin{figure}[p]
\hbox to 1.5cm {\hfil\mbox{
$d\sigma^{cc}_{\pi^- p}/d y$, 
 $d\sigma^{J/\Psi+D\bar D}_{\pi^- p}/d y$, pb}}
\epsfxsize=14cm \epsfbox{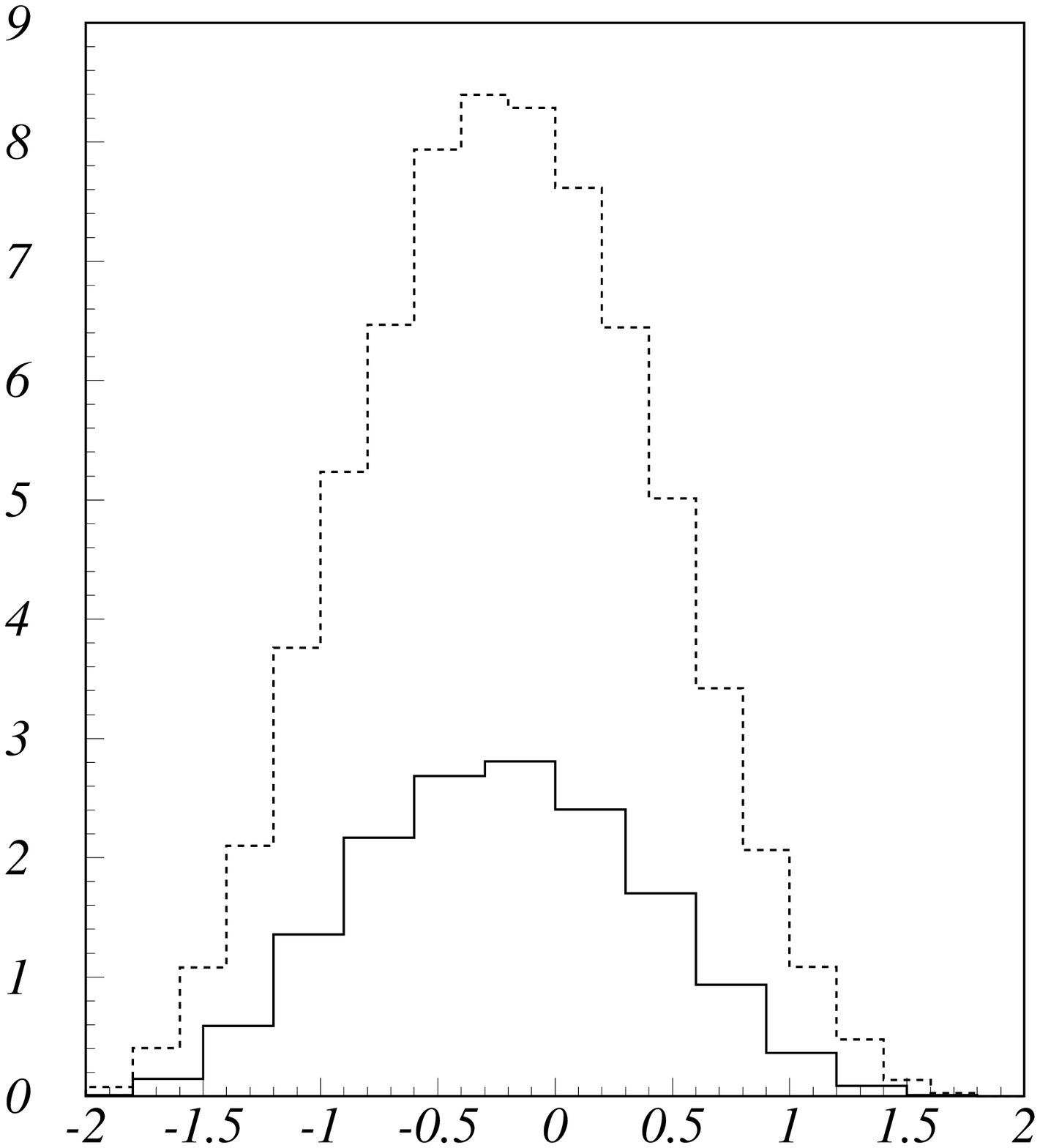}

\vspace*{-1.cm}
\hbox to 14.cm {\hfil \mbox{$y$}}
\vspace*{1.cm}
\caption{
$d\sigma^{cc}_{\pi^- p}/d y$ (solid histogram)
£ $d\sigma^{J/\Psi+D\bar D}_{\pi^- p}/d y$ (dashed histogram)
at the pion-proton interaction energy of 35 GeV.}
\label{fig5b}
\end{figure}

\begin{figure}[p]
\hbox to 1.5cm {\hfil\mbox{
$d\sigma^{cc}_{pp}/d p_T$, 
 $d\sigma^{J/\Psi+D \bar D}_{pp}/d p_T$, pb/GeV}}
\epsfxsize=14cm \epsfbox{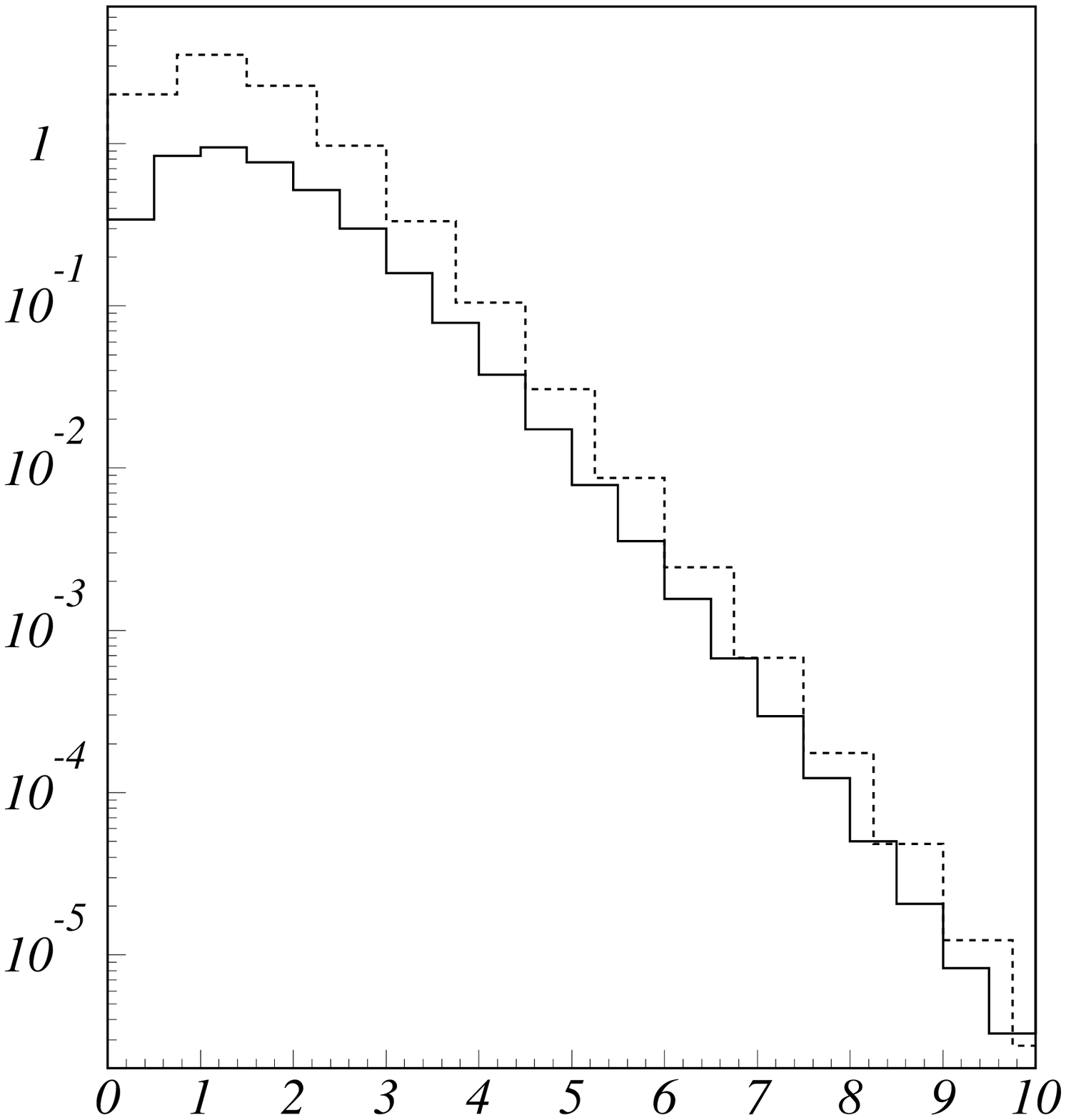}

\vspace*{-1.cm}
\hbox to 14.cm {\hfil \mbox{$p_T$, GeV}}
\vspace*{1.cm}
\caption{
$d\sigma^{cc}_{pp}/d p_T$ (solid histogram)
and $d\sigma^{J/\Psi+D\bar D}_{pp}/d p_T$ (dashed histogram)
at the proton-proton interaction energy of 40 GeV.}
\label{fig6a}
\end{figure}

\begin{figure}[p]
\hbox to 1.5cm {\hfil\mbox{
$d\sigma^{cc}_{pp}/d y$, 
 $d\sigma^{J/\Psi+D\bar D}_{pp}/d y$, pb}}
\epsfxsize=14cm \epsfbox{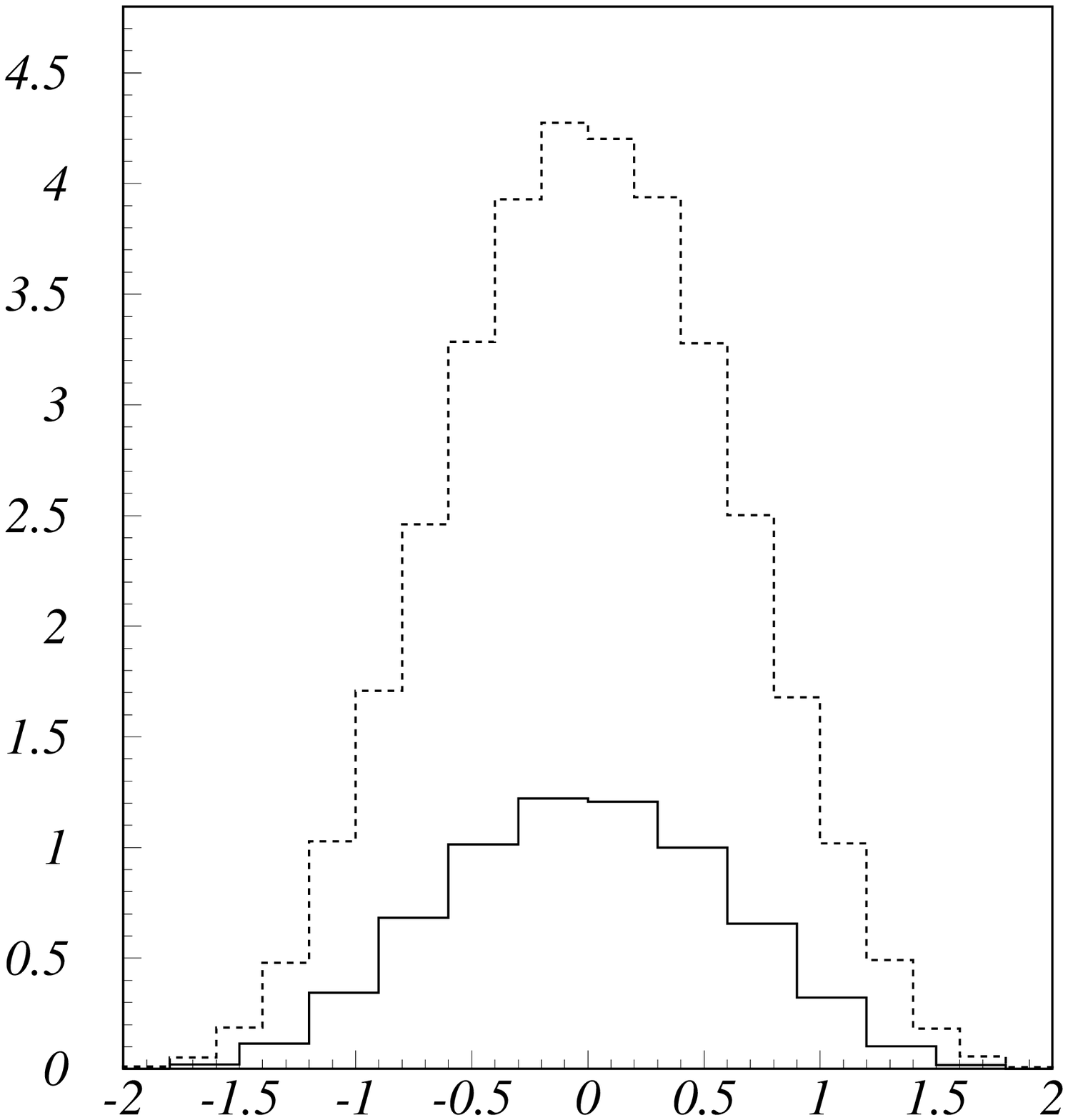}

\vspace*{-1.cm}
\hbox to 14.cm {\hfil \mbox{$y$}}
\vspace*{1.cm}
\caption{
$d\sigma^{cc}_{\pi^- p}/d y$ (solid histogram)
£ $d\sigma^{J/\Psi+D\bar D}_{\pi^- p}/d y$ (dashed histogram)
at the proton-proton interaction energy of 40 GeV.}
\label{fig6b}
\end{figure}

The ratio for the $(cc)$-diquark production and total
charm production is
$\sigma_{(cc)}/\sigma_{charm} \sim 10^{-4} -10^{-3}$  in  the
collider
experiments and $\sim 10^{-6} -10^{-5}$ in the fixed target
experiments.
The same situation is observed for the hadronic $J/\Psi+D \bar D$
production.
The distributions for the $(ccq)$-baryon and $J/\Psi+D \bar D$
production are shown in Figs. \ref{fig5a}--\ref{fig6b} for the
$\pi^-p$-interaction
at 35 GeV and for the $pp$-interaction at 40 GeV, correspondingly.
The rapidity distributions in Figs. \ref{fig5b} and \ref{fig6b} point
to the central
state of $(ccq)$-baryon production and that for 
$J/\Psi +D\bar D$.

The $p_T$-distributions  of  these processes are alike to each other
also (we assume that at the given energies the $(cc)$-diquark has
no fragmentational transition into the  baryon,
but it catch up the light quark from the quark-antiquark pair sea).
One can
see in the latter Figs., that the process of $J/\Psi +D\bar D$
production
can be used to normalize the estimate of $(ccq)$-baryon yield,
wherein
the following additional uncertainties appear as
\begin{enumerate}
\item
the unknown value of $|R_{(cc)}(0)|^2$,
\item
uncertainties related with the hadronization of $(cc)$-diquark.
\end{enumerate}

One can see from the given estimates that in the experiments  with
the expected number of charmed events at the level about $10^6$ 
(for example, in the E781 experiment, where $\sqrt{s}=35$ GeV), 
one has to expect about one
event with the doubly charmed baryon.  The situation is more
promising and
pleasant for the
$pp$-interaction at 800 GeV (HERA-B). The considered processes yield
about $10^5\; \Xi_{cc}^{(*)}$-baryons and a close number  of
$J/\Psi +D\bar D$ in the experiment specialized for the detection of
about $10^8$ events with the $b$-quarks.

\section{Production of (ccq)-baryon at Colliders}
As one can see in the previous Section, the observation of
$\Xi_{cc}^{(*)}$-baryons  presents a rather difficult problem
in the experiments specialized for the study of
charmed particles.  As a rule, such experiments are 
carried out at fixed targets, so that the effective value of
subprocess energy is strongly decreased. So, 
the relative contribution of doubly
charmed baryons into the total charm yield is of the order of
$10^{-6}-10^{-5}$. The production of $(ccq)$-baryons at colliders 
with large $p_T$ is more effective.
In this case the cross-section is determined by the region of
quark-antiquark and gluon-gluonic energy, where the threshold effect
becomes negligible  and the partonic luminosities are quite large
at $x\sim M/\sqrt{s}$. So, the suppression factor in respect to the
single
production of $c\bar c$-pairs is much less and it is in the range of
$10^{-4}-10^{-3}$.

\begin{figure}[p]
\hbox to 1.5cm {\hfil\mbox{
$d\sigma^{\Xi_{cc}^{(*)}}_{pp}/d p_T$, 
 $d\sigma^{J/\Psi+D\bar D}_{pp}/d p_T$, nb/GeV}}
\epsfxsize=14cm \epsfbox{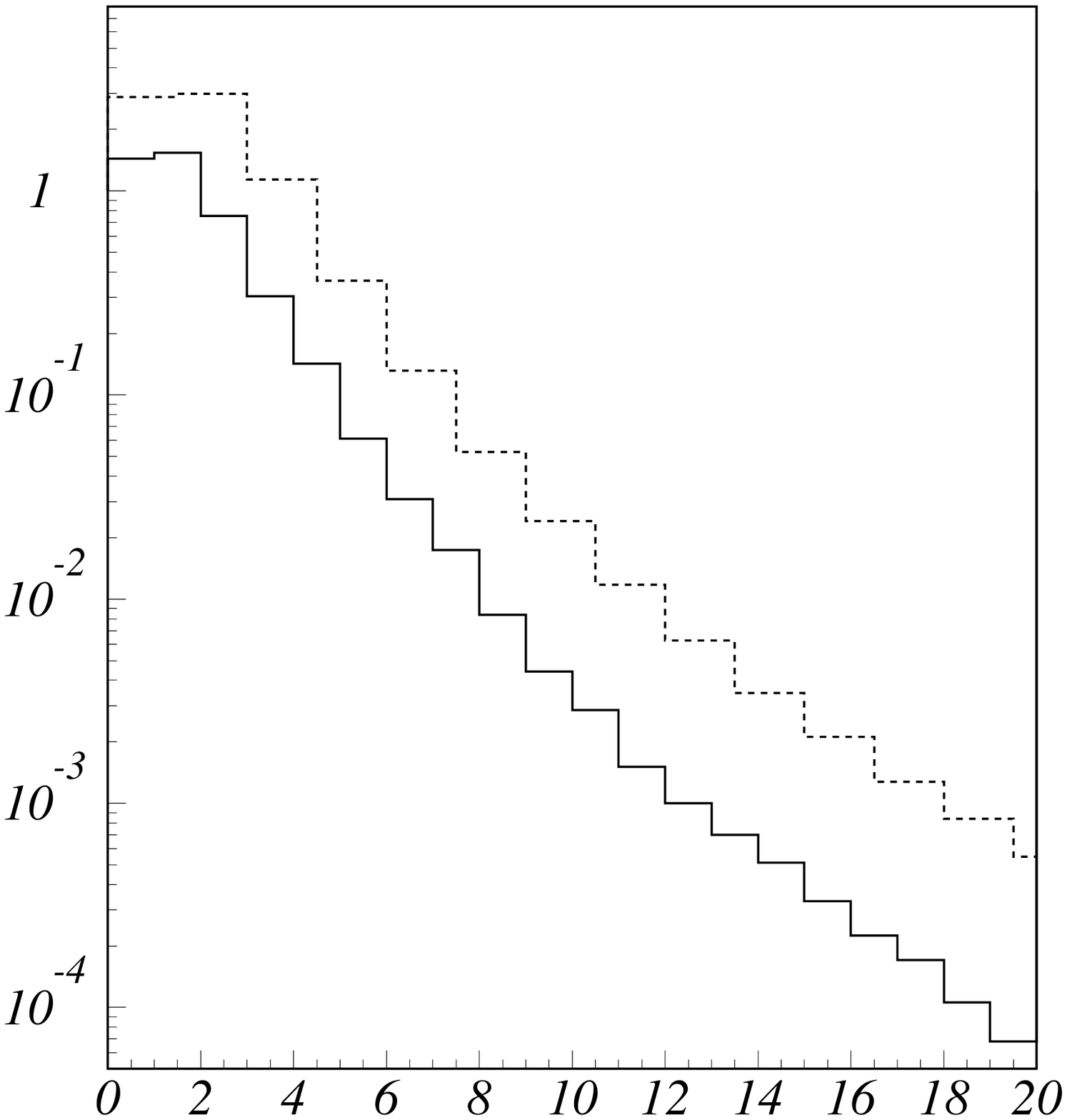}

\vspace*{-1.cm}
\hbox to 14.cm {\hfil \mbox{$p_T$, GeV}}
\vspace*{1.cm}
\caption{
$d\sigma^{\Xi_{cc}^{(*)}}_{pp}/d p_T$ with taking into
account the fragmentation of $cc$-diquark into
$\Xi_{cc}^{(*)}$-baryon (solid histogram)
and  $d\sigma^{J/\Psi+D\bar D}_{pp}/d p_T$ (dashed histogram)
at proton-proton interaction energy of 1.8 TeV.}
\label{fig7a}
\end{figure}

\begin{figure}[p]
\hbox to 1.5cm {\hfil\mbox{
$d\sigma^{\Xi_{cc}^{(*)}}_{pp}/d p_T$, 
 $d\sigma^{J/\Psi+D\bar D}_{pp}/d p_T$, nb/GeV}}
\epsfxsize=14cm \epsfbox{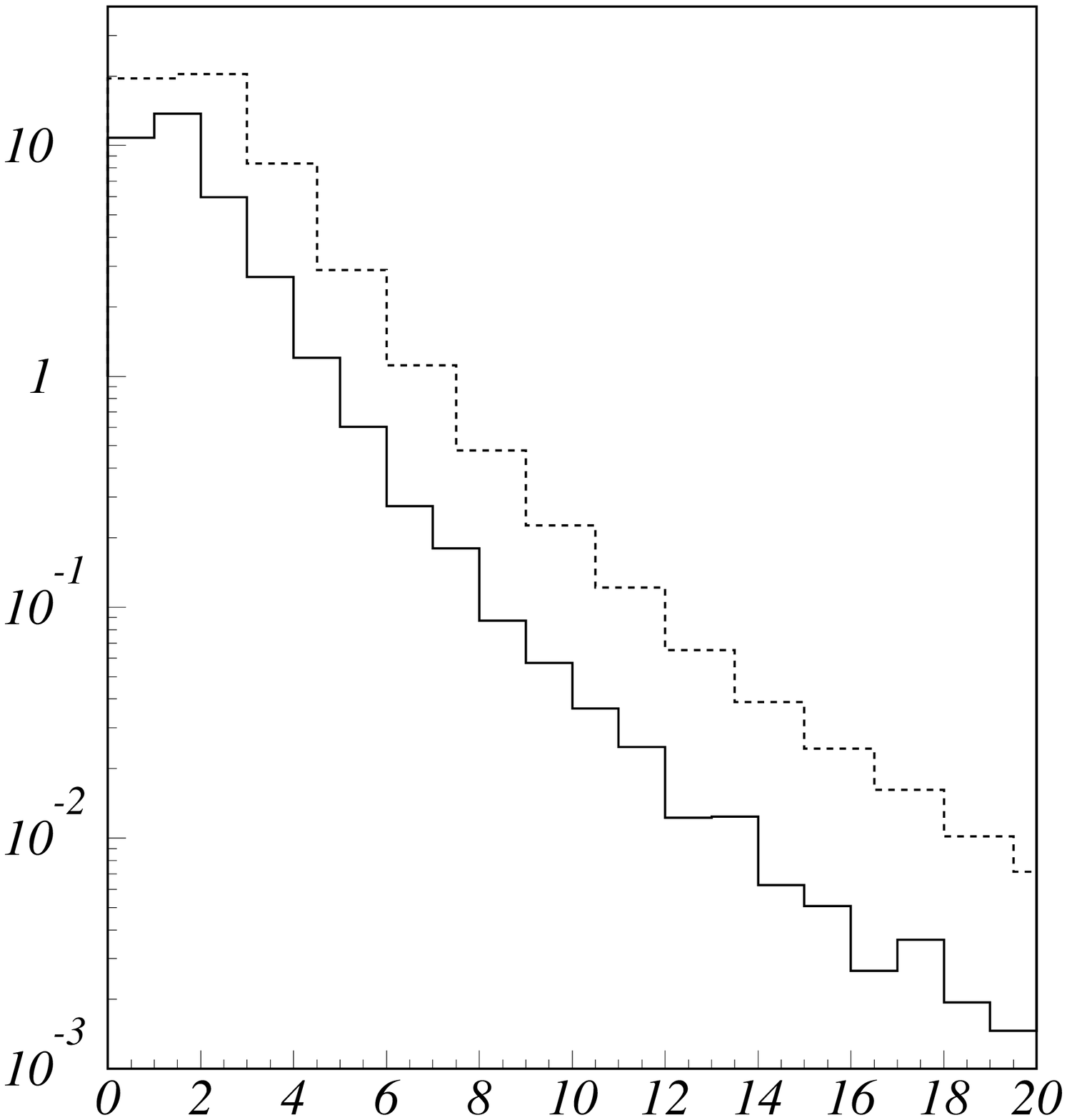}

\vspace*{-1.cm}
\hbox to 14.cm {\hfil \mbox{$p_T$, GeV}}
\vspace*{1.cm}
\caption{
$d\sigma^{\Xi_{cc}^{(*)}}_{pp}/d p_T$ with taking into
account the fragmentation of $cc$-diquark into
$\Xi_{cc}^{(*)}$-baryon (solid histogram)
and  $d\sigma^{J/\Psi+D\bar D}_{pp}/d p_T$ (dashed histogram)
at proton-proton interaction energy of 14 TeV.}
\label{fig7b}
\end{figure}

 The $p_T$-distributions for $\Xi_{cc}^{(*)}$
and $J/\Psi$ (which is produced with $D$ and $\bar D$) at TEVATRON
and
LHC are shown in Figs. \ref{fig7a} and \ref{fig7b}. 
The rapidity cut ($|y|<1$) is taken into account.

One can easily understand that
the presented $\Xi_{cc}^{(*)}$ cross-sections are
the upper estimates for the real cross-sections because of the
possible
dissociation of  heavy diquark into the $D D$-pair.

Further,
even if the $(cc)$-diquark, being the color object, transforms into
the baryon
with the unit probability, one has to introduce 
the fragmentation function describing  the hadronization of
diquark into the baryon at quite large $p_T$ values. The simplest
form of this
function can be chosen by analogy with that for the heavy quark
\begin{equation}
D(z)\sim
\frac{1}{z}\frac{1}{(m^2_{cc}-\frac{M^2}{z}-\frac{m_q^2}{1-z})^2},
\label{d}
\end{equation}
where $M$ is the mass of  baryon $\Xi_{cc}^{(*)}$,
$m_{cc}$ is the mass of diquark, $m_q$ is the mass of light quark
(we suppose it to be equal to 300 MeV).

The $p_T$-distributions of doubly charmed baryon production, as those
are shown 
in Figs.  \ref{fig7a} and \ref{fig7b}, are calculated with the use of
(\ref{d}).

One has to mention, that in the leading order over the inverse heavy 
quark mass, the relative yield of $\Xi_{cc}$ and $\Xi_{cc}^*$
is determined by the
simple counting rule for the spin states, and it equals
$\sigma (\Xi_{cc}):\sigma (\Xi_{cc}^*)=1:2$.
In this approach one does not take into account a possible difference
between the fragmentation functions for the baryons with the
different spins. The corresponding difference is observed in
the perturbative fragmentation functions for
the heavy mesons and quarkonia \cite{12}.

\section{Discussion}

As we have shown on the basic of perturbative calculations
for the hard production of doubly charmed diquark fragmentating in
the
baryon, the observation of doubly charmed
baryons is a difficult problem, because the ratio of
$\sigma ( \Xi_{cc}^{(*)}) /\sigma ( charm)$ for these baryons and
charmed
particles yields the value of $10^{-6}-10^{-3}$ depending on the
process
energy.
The suppression of doubly charmed baryon yield at low energies  is
explained by threshold effect. As one can see in accordance with 
Tab. 1, about $10^5$ events 
with the production of $\Xi_{cc}^{(*)}$-baryons
can be expected at HERA-B. Practically  the same  number of events at
$p_T>5$ GeV and  $|y|<1$ is expected at TEVATRON with
the integrated luminosity of $100\ {\rm pb}^{-1}$. The large
luminosity
and large interaction energy allow one to increase the yield of the
doubly
charmed baryon by $10^4$ times at LHC.

Under conditions of large yield of the doubly charmed baryons, the
problem
of their registration appears.

First of all it is interesting to estimate the lifetimes of the
lightest
states of $\Xi_{cc}^{++}$ and $\Xi_{cc}^{+}$. The simple study of
quark diagrams shows that in the decay of  $\Xi_{cc}^{++}$-baryons
the Pauli interference
for the decay products of charmed quark and valent quark
in the initial state
takes place as well as in the case of $D^+$-meson decay.
In the decay of $\Xi_{cc}^+$ the exchange by the $W$-boson  between
the
valent quarks plays an important role as well as in the decay of
$D^0$.
Therefore we suppose that the  mentioned mechanisms give
the same ratio for the both baryon and $D$-meson
lifetimes
$$
\tau(\Xi_{cc}^+)\approx 0.4\cdot \tau(\Xi_{cc}^{++}).
$$

The presence of two charmed quark in the initial state
results in following expressions
\begin{eqnarray}
\tau(\Xi_{cc}^{++})&\approx & \frac{1}{2} \tau(D^+)\simeq 0.53\; {\rm
ps,}
\nonumber \\
\tau(\Xi_{cc}^{+})&\approx & \frac{1}{2} \tau(D^0)\simeq 0.21\;{\rm
ps.}
\nonumber
\end{eqnarray}

One can point to the important decay modes of these baryons in
analogy with
the case of the charmed mesons
$$
{\rm BR}(\Xi_{cc}^{++}\to K^{0(*)}\Sigma_c^{++(*)})\approx
{\rm BR}(\Xi_{cc}^{+}\to
K^{0(*)}(\Sigma_c^{+(*)}+\Lambda_c^+))\approx
{\rm BR}(\Lambda_c\to K^{0(*)}p)\simeq 4\cdot 10^{-2}.
$$

One can observe $4\cdot 10^{3}$ events in these decay modes at HERA-B
and
TEVATRON without taking into account a detection efficiency. One has
to expect the yield of $4\cdot 10^7$ such decays at LHC.
Among other decay modes, $\Xi_{cc}^{++}\to \pi^+\Xi_c^+$  and
 $\Xi_{cc}^{+}\to \pi^+\Xi_c^0$ taking place with the probability of
 about
1\%, can be essential.

The excited $\Xi_{cc}^*$ states always
decay into $\Xi_{cc}$ by the emission of
$\gamma$-quanta so that the branching fraction of transition is
equal to 100\%, since the emission of $\pi$-meson is impossible in
the
$\Xi_{cc}^*$ decay  because
of the small value of splitting between the basic state and the
excited one,
in contrast to the charmed  meson decay.

In conclusion we mention another possibility to increase the yield of
doubly charmed baryons in fixed  target experiments. In the model
of  intrinsic charm \cite{13} one assumes, that the nonperturbative
admixture of exotic hybrid state $|c\bar c uud \rangle$ presents in
the proton
along with the ordinary state $|uud\rangle$
including three light valent quarks.
The probability $P_{ic}$ of $|c\bar c uud\rangle$-state is suppressed
at the level of 1\%.  The valent charmed quark from that state can
recombinate with the charmed quark produced in the hard partonic
process of the $(c\bar c)$-pair production. The energy dependence for
such
doubly charmed baryon production repeats one for the single
charmed quark production
in the  framework of pQCD up to the factor of exotic state
suppression and 
the factor of fusion of two charmed quarks into the diquark, $K\sim
0.1$.
This mechanism has no threshold of four quark state production in
contrast to the discussed perturbative one. Therefore
at low energies of fixed target experiments, where the threshold
suppression of perturbative
mechanism is strong, the model of intrinsic
charm would yield the dominant contribution into the
$\Xi_{cc}^{(*)}$ production.
So, the number of events in this model would be increased by three
orders of
magnitude, and ratio of  $\Xi_{cc}^{(*)}$ and  charmed particle
yields would
equal $\sigma(\Xi_{cc}^{(*)})/\sigma(\rm{charm})\sim 10^{-3}$.
At high energies the perturbative production is comparable  with the
intrinsic charm contribution.  One has to note, that the
$|c\bar c c\bar c uud\rangle$-state suppressed
at the level of $3\cdot 10^{-4}$,
also could increase the doubly charmed baryon production at low
energies of hadron-hadron collisions.

Thus, the observation of $\Xi_{cc}^*$-baryons in hadronic
interactions is
a quite realistic problem, whose solution
opens new possibilities to research the heavy
quark interactions. The observation
of $\Xi_{cc}^{(*)}$-baryons at fixed target experiments \cite{14}
would allow one to investigate the contributions of different
mechanisms
in the doubly charmed baryon production, as the contribution of the
perturbative mechanism and that of the intrinsic charm, which
strongly increases the yield of these baryons.

The authors express their gratitude to A.Kulyavtsev for the
discussion
of problem under consideration.

This work is supported,  in part, by the Russian Foundation for
Basic Research, grants 96-02-18216 and 96-1596575. The work of A.V. Berezhnoy
has been made possible by a fellowship of INTAS Grant 93-2492
and is carried out within the research program of International
Center for Fundamental Physics in Moscow.

\newpage
\begin{table}[p]
\caption{The production cross-section of doubly charmed baryons
at different facilities.}
\begin{tabular}{|c|c|c|c|c|}
facility & HERA-B & E781 & TEVATRON & LHC\\
\hline
\parbox{3.5cm}{ \vspace*{1mm} total cross-section, nb/nucleon}
 & $2.011\cdot 10^{-3}$ & $4.582\cdot 10^{-3}$
 & $11.61$ & $122$\\
\end{tabular}
\end{table}

\newpage
\newpage
\newpage
\newpage

\newpage
\newpage
\newpage
\newpage
\newpage
\newpage
\newpage
\newpage
\end{document}